\documentclass[aps,prl,preprint,showpacs,superscriptaddress]{revtex4}

\usepackage{graphicx,amssymb}

\begin{document}

\title{Motion of a sphere through an aging system}

\author{H. Tabuteau}

\affiliation{Department of Physics and Astronomy, University of
Western Ontario, London ON Canada N6A 3K7}

\author{John R. de Bruyn}

\affiliation{Department of Physics and Astronomy, University of
Western Ontario, London ON Canada N6A 3K7}

\author{P. Coussot}

\affiliation{Department of Physics and Astronomy, University of
Western Ontario, London ON Canada N6A 3K7}

\affiliation{Institut Navier, Paris, France }

\date{\today}

\begin{abstract}

We have investigated the drag on a sphere falling through a clay
suspension that has a yield stress and exhibits rheological aging.
The drag force increases with both speed and the rest time between
preparation of the system and the start of the experiment, but
there exists a nonzero minimum speed below which steady motion is
not possible. We find that only a very thin layer of material
around the sphere is fluidized when it moves, while the rest of
suspension is deformed elastically. This is in marked contrast to
what is found for yield-stress fluids that do not age.

\end{abstract}

 \maketitle

Pasty materials and concentrated suspensions have microscopic
internal structure which gives them the ability to resist shear.
Consequently, they can behave as soft solids or shear-thinning
fluids, depending on the stress applied \cite{c05}. When the force
$F$ acting on an object within the material is less than a
critical value $F_c$, the material responds as a solid and the
object will not move. When $F > F_c$, the material around the
object becomes fluidized and the object moves \cite{btab85,tcd06}.
Pasty materials typically exhibit rheological aging
\cite{sollich}, that is, their rheological properties change with
time due to slow evolution of the microstructure. The dynamics of
this evolution is similar to that of glasses \cite{sollich,d00},
and in some cases these materials can undergo a fluid-solid
transition as they age. They can also easily develop strain
heterogeneities such as shear banding \cite{c02}.

The motion of particles through pasty materials is important in
many applications \cite{c05}, and drag has been studied to some
extent in yield-stress systems that do not age
\cite{acu95,bm97,btab85,tcd06}. The drag force on a sphere of
radius $R$ moving through a yield-stress fluid is given by
\begin{equation}
\label{integral} F = -\int_{S_c}\left(\textbf{T}\cdot \hat e
\right) \cdot \hat x \;ds.
\end{equation}
$\textbf{T}$ is the stress tensor and $S_c$ the material surface
on which the yielding criterion is met \cite{c05}; $S_c$ separates
the solid and liquid regions in the material surrounding the
sphere. $\hat e$ is the outward-pointing unit vector at $S_c$ and
$\hat x$ a unit vector in the direction of the motion. In simple
shear, the stress $\tau$ and shear rate $\dot\gamma$
\textit{without} aging can be related by the Herschel-Bulkley
model, $\tau/\tau_c = 1+\left(K/\tau_c\right) \dot\gamma^n$, where
$\tau_c$ is the yield stress and $K$ and $n$ are material
parameters. The drag force can be written similarly \cite{bm97}:
\begin{equation}
\label{force} F/F_c = 1+\left(K/\tau_c\right) \dot\gamma_{app}^n
\end{equation}
for $F > F_c$.  $F_c$, the minimum force required for steady
motion in the liquid regime, has been calculated \cite{btab85} to
be
\begin{equation}
\label{yield_force} F_c=4\pi R^2 \tau_c k_c
\end{equation}
with $k_c=3.5$. The apparent shear rate $\dot\gamma_{app} =
v/\ell$, where $v$ is the sphere's speed and $\ell = 1.35R$ for
$n=0.5$ \cite{bm97}. With these values for $k_c$ and $\ell$, Eq.
(\ref{integral}) indicates that, in the absence of aging, $S_c$ is
approximately $R$ from the sphere's surface. Experiments with a
non-aging yield-stress fluid agreed well with these theoretical
results \cite{tcd06}. The effect of aging on drag has received
little attention, however  \cite{f04}, and remains poorly
understood from both the macroscopic and microscopic points of
view.

Here we study the drag on a weighted ping-pong ball of diameter
$2R= 3.96$ cm falling through a clay suspension that exhibits
rheological aging. Since the sphere falls through previously
undisturbed material, its motion at a time $t$ probes the
properties of the suspension at that time and provides a direct
measure of the changes in properties as the suspension ages. We
find that the drag force increases with both $v$ and the waiting
time $t_w$ between preparation of the sample and the start of the
experiment, and that $v$ approaches a non-zero value as $F$
approaches the critical force $F_c$. We show that the sphere moves
through the material by fluidizing a very thin layer of the
suspension --- much smaller than $R$ --- immediately surrounding
it, while the rest of suspension is deformed elastically. This is
very different from the situation in non-aging materials
\cite{tcd06}.

We worked with Laponite RD \cite{laponite}, a synthetic clay known
to exhibit aging \cite{bkhlm03}. Laponite consists of disk-shaped
particles 30 nm in diameter and 1 nm thick. We mixed the clay 3\%
by weight with deionized water, adding NaOH to raise the pH to 10.
The resulting suspension, with density $\rho_l = 1012$ kg/m$^3$,
was then stored in a sealed container for four weeks to ensure
complete hydration of the clay particles. Thorough remixing before
each experiment was critical for obtaining reproducible results. A
combination of local and large-scale mixing effectively destroyed
the suspension's microstructure and put it in a reproducible
initial state. After remixing the material was left to age for a
waiting time $t_w$, during which its microscopic structure
partially reformed, then the experiment was started.

The ping-pong ball had a small hole cut into it, and its density
$\rho_s$ was changed by gluing small steel beads inside it. Apart
from a small piece of tape at the top to cover the hole, its
surface roughness was on the order of 10 $\mu$m. This is much
larger than the particle size, so slip at the sphere's surface is
not expected to be important. It was dropped into the suspension
from 1 cm above the free surface and its position recorded at up
to 250 frames per second. The experimental container was square in
cross-section, with width $L = 20$ cm and depth  45 cm. For our
value of $2R/L = 0.2$, wall effects are small \cite{acu95}.

\begin{figure}[htbp]
\includegraphics[width=6in]{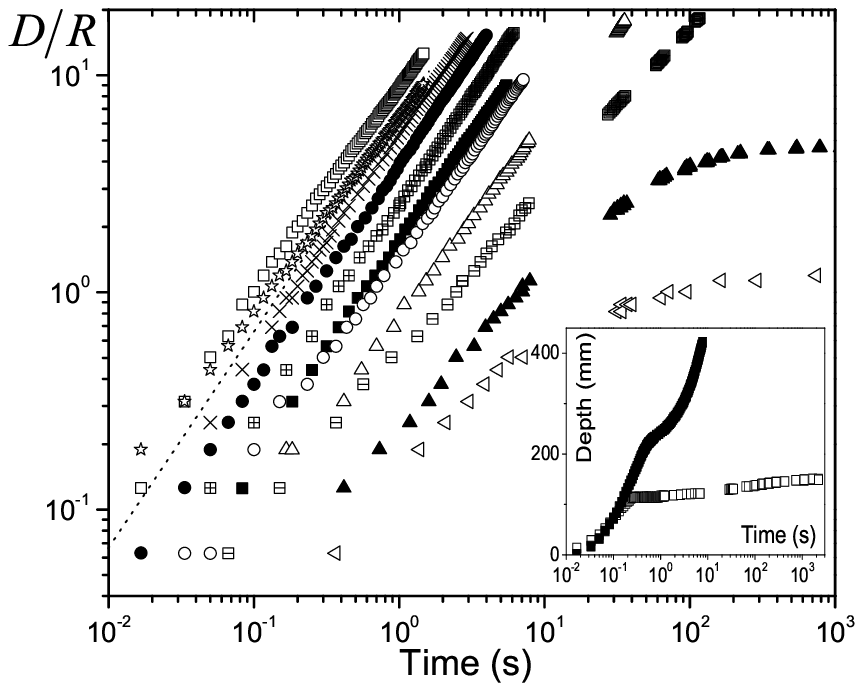}
\caption{\label{vsdeltarho} Scaled penetration depth $D/R$ vs.
time $t$ for the sphere falling through the Laponite suspension
for, from left to right, $\Delta\rho = 1589$, 1524, 1460, 1389,
1319, 1251, 1153, 1057, 956, and 890 kg/m$^3$. Here $t_w = 4$ min.
The dotted line has a logarithmic slope of 1, corresponding to a
constant speed. The inset shows $D$ as a function of $t$ for $t_w
= 20$ min and $ \Delta\rho = 1806 $ kg/m$^3$ (solid squares) and
1151 kg/m$^3$ (open squares).}
\end{figure}

\begin{figure}[htbp]
\includegraphics[width=6in]{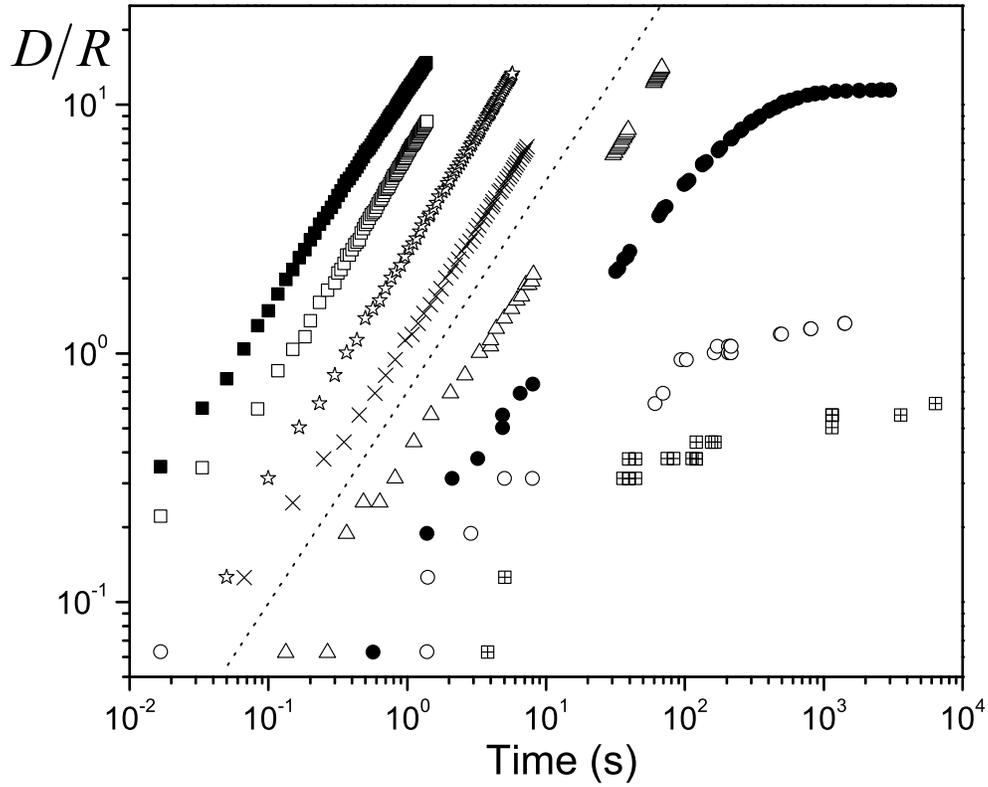}
\caption{\label{vstw} $D/R$ as a function of $t$ for $\Delta\rho =
1345$ kg/m$^3$ and, from left to right, $t_w = 1$, 2, 5, 10, 20,
30, 40, and 45 min. The dotted line has a logarithmic slope of 1.
}
\end{figure}

After an early-time regime dominated by inertia, the sphere either
slows to a steady terminal velocity or stops, as shown in the
inset to Fig. \ref{vsdeltarho}. We focus on later times, when
inertial effects can be neglected. Figure \ref{vsdeltarho} shows
the sphere's penetration depth $D$ (scaled by $R$) as a function
of time $t$ for different values of $\rho_s$ and a fixed $t_w$
\cite{origin}. We can distinguish three regimes \cite{f04},
depending on the net gravitational force $F \propto \Delta\rho =
\rho_s-\rho_l$ on the falling sphere. For sufficiently large
$\Delta\rho$ (Regime 1), the sphere reaches a steady-state speed
as gravity becomes balanced by drag. For small $\Delta\rho$
(Regime 3), the speed of the sphere decreases to zero while the
sphere moves a distance of order $R$. In the intermediate regime
(Regime 2), the initial velocity $v_i$ persists up to a
displacement much larger than $R$, after which the sphere again
slows and stops. The same three regimes are observed as $t_w$ is
increased for fixed $\Delta\rho$, as shown in Fig. \ref{vstw}.

Fig. \ref{vstw} shows that $v_i$ decreases with increasing $t_w$
for a fixed $\Delta\rho$. The instantaneous speed $v$ also
decreases with $t$ in Regimes 2 and 3, as the material continues
to age. This tempts one to describe the material as having an
apparent viscosity which increases with time as structure in the
suspension redevelops. This naive model is inconsistent with the
data, however: For both of the two largest $t_w$ values plotted in
Fig. \ref{vstw}, the sphere starts to slow significantly after
about 100 s, much shorter than the 5 min difference in $t_w$
between these two trials. If the stoppage were due to an increase
in viscosity, this viscosity would have to increase much more over
the 100 s of motion than over the additional 5 min of aging. This
contradiction shows that a closer look is required.

Our results in fact indicate that the state of the material and
the flow around the moving sphere are very different in the three
regimes. The overall deformation $\gamma$ of the suspension in
Regime 3 is of order $2R/L \approx 0.2$, typical of the critical
deformation $\gamma_c$ at which pasty materials undergo the
transition from solid to liquid \cite{c06}. This suggests that our
suspension remains a solid in Regime 3. In Regime 1, in contrast,
$\gamma \gg \gamma_c$ and the material around the sphere is
quickly fluidized. It is also initially fluidized in Regime 2, but
here the material structure redevelops on a time scale similar to
that of the flow, leading to a transition from liquid to solid
over the course of the run. This description is consistent with,
for example, the data for $t_w=30$ (in Regime 2) and 40 min (in
Regime 3). In the former case, the sphere stops after about 600 s
because the material has changed from liquid to solid, while in
the latter case the material is solid at the start of the run. The
material becomes solid at about the same age --- approximately 40
min --- in both cases. In Regimes 1 and 2, $v_i$ depends on the
state of development of the microstructure, and so on $t_w$. $v_i$
is not zero in Regime 3, however, as seen in Figs.
\ref{vsdeltarho} and \ref{vstw}, so there is a minimum speed
$v_c\sim 1$ mm/s that can be reached in the liquid regime. If $v_i
< v_c$, the material is in its solid regime, steady motion of the
sphere is not possible, and it comes quickly to a stop
\cite{steady}.

\begin{figure}[htbp]
\includegraphics[width=6in]{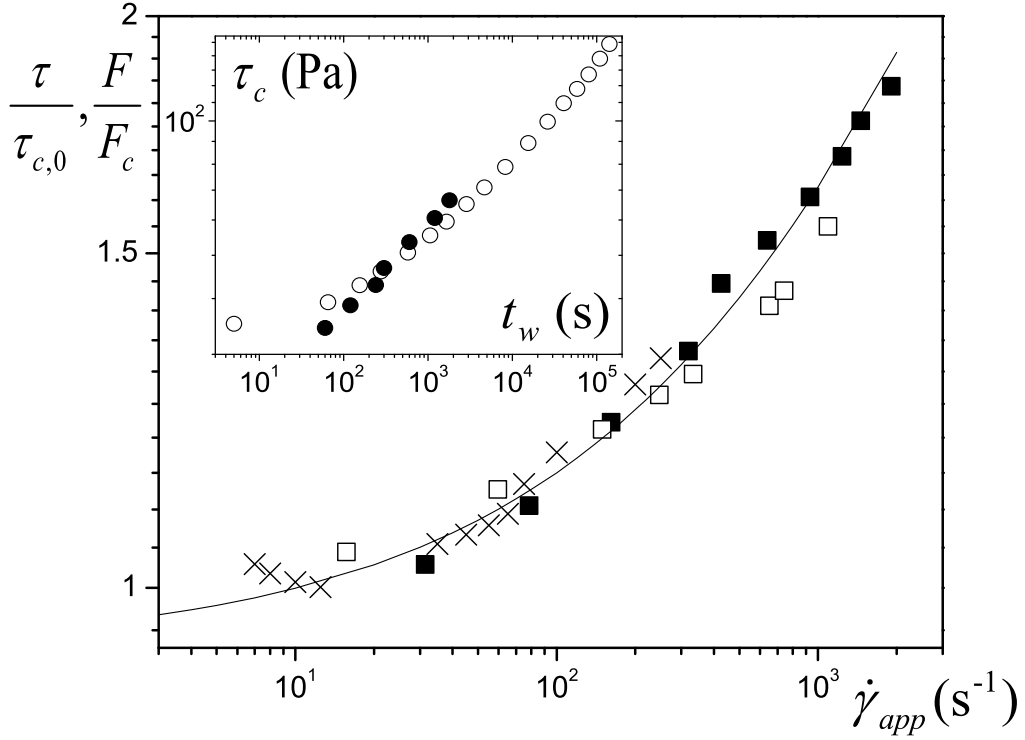}
\caption{\label{dimensionless} Crosses: $\tau/\tau_{c,0}$ as a
function of $\dot\gamma_{app}$ for the Laponite suspension in
simple shear. Squares: $F/F_c$ vs. $v/\ell$ obtained from the
falling-sphere experiments with $t_w= 4$ min (solid squares) and
20 min (open squares). The curve is a fit of Eq.
\protect\ref{foverfc} to the force data. The inset shows the
increase of the apparent yield stress $\tau_c$ with age determined
from $G'(t_w)$ (open circles) and $F_c(t_w)$ (solid circles). The
values of $k_c$ and $\ell$ used in this analysis are discussed in
the text. }
\end{figure}

The rheological properties of our suspension were measured with an
ARES RHS controlled-strain rheometer, using a 50 mm parallel-plate
tool (plate separation 1.5 mm) covered with sandpaper to prevent
slip. Following a preshear to rejuvenate the sample, we measured
the stress $\tau$ as a function of $\dot\gamma$. We started from
high $\dot \gamma$ and worked downwards to limit the effects of
aging on the measurements, and waited 15 s at each value of
$\dot\gamma$. The resulting flow curve, shown by the crosses in
Fig. \ref{dimensionless}, has a minimum at $\dot\gamma_c \approx
20$ s$^{-1}$, possibly indicating the presence of shear-banding at
lower $\dot \gamma$ \cite{pmp96}. We take only the data for $\dot
\gamma > \dot \gamma_c$ as reflecting the behavior of the
homogeneous material. The instability of uniform shear flow below
$\gamma_c$ may be analogous to our observation that steady motion
of the falling sphere is not possible for $v < v_c$, suggesting
that flow instabilities analogous to shear-banding are not
restricted to simple viscometric flows.

The elastic modulus $G'(t_w)$ was measured by applying an
oscillatory shear with angular frequency $\omega = 1$ rad/s and
deformation amplitude 5\%. $G'$ was independent of $\omega$ for
$0.1 <\omega < 100$ rad/s. We found $G'$ to increase significantly
with $t_w$. Taking the material to be viscoelastic in the solid
regime, the yield stress is given by $\tau_c = G'\gamma_c$. Using
the $t_w= 0$ limit of $G'$ and $\gamma_c = 0.2$, we find
$\tau_{c,0}=34$ Pa, consistent with the value estimated from the
minimum in the flow curve. The stress data in Fig.
\ref{dimensionless} have been normalized by this value.
$\tau_c(t_w)$ was estimated from $G'(t_w)$ in the same way and is
plotted in the inset to Fig. \ref{dimensionless}. The yield stress
increases smoothly with $t_w$ with a logarithmic slope that
increases as the material ages.

We can describe the net force $F$ on our sphere by
\begin{equation}\label{foverfc}
F/F_c(t_w) = a + b \dot\gamma_{app}^n
\end{equation}
for $F > F_c$. This is similar to Eq. (\ref{force}) \cite{acu95},
but here $F_c$ depends on $t_w$ and $a$ and $b$ are treated as
parameters. A fit to our falling-sphere data gives $a = 0.93$,
less than the value of 1 expected for non-aging yield-stress
fluids. Thus the shear rate, and so the speed $v$ of the sphere,
is greater than zero when $F = F_c$, as discussed above. $F/F_c$
is plotted as a function of $\dot\gamma_{app}$ for two values of
$t_w$ in Fig. \ref{dimensionless}. To make these data consistent
with the rheometric flow curve, we had to take $\ell = 0.009R$ in
the calculation of  $\dot\gamma_{app}$, a factor of 100 smaller
than for non-aging yield-stress fluids. This suggests that the
sheared region around the moving sphere is very thin for our aging
material.

Using Eq. (\ref{yield_force}) and the data in Fig. \ref{vstw}, we
can determine $F_c(t_w)$ for a given $\Delta\rho$. We then
calculate the corresponding yield stress using Eq.
(\ref{yield_force}). We again find that the yield stress increases
with $t_w$, but to get agreement between the values of $\tau_c$
determined from the falling-sphere data and those obtained from
the measurements of $G'$, we are forced to take $k_c = 1.085$ in
Eq. (\ref{yield_force}). This is much less than the value for
non-aging yield-stress fluids, and close to what is calculated
from Eq. (\ref{integral}) if $S_c$ coincides with the surface of
the sphere \cite{da67}. This is further evidence that the moving
sphere fluidizes only a thin layer of the suspension, with the
lowness of $k_c$ resulting from the fact that less force is
required to disrupt the structure of a smaller volume of material.
We emphasize that the low values of $k_c$ and $\ell$ do not depend
on any choice of constitutive relation, but are required to make
the falling-sphere data consistent with the rheometric
measurements.

\begin{figure}[htbp]
\includegraphics[width=6in]{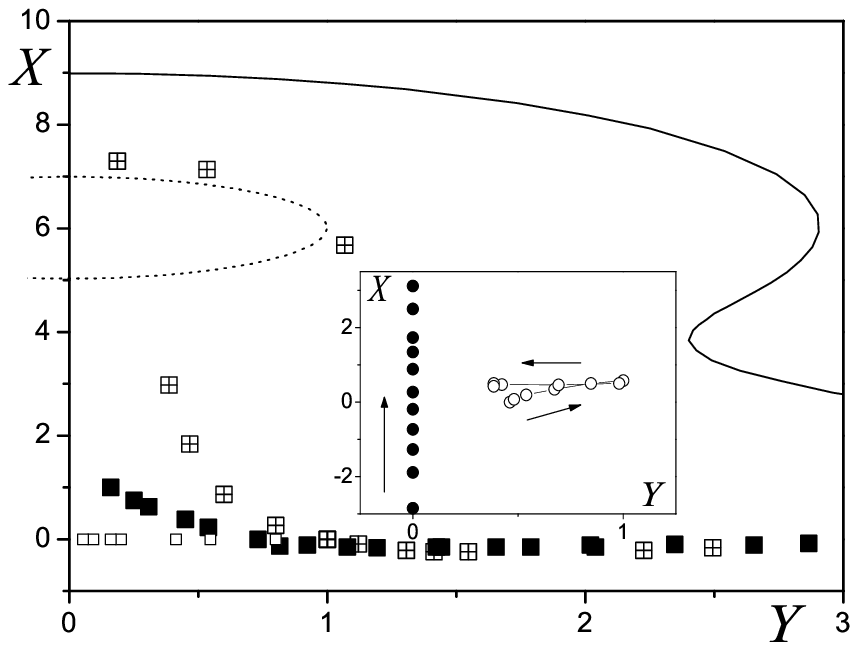}
\caption{\label{displacement} Displacement $X$ of a line of tracer
particles initially at $X =0$ after the sphere moves from $X=-3$
to $X=6$, plotted against the distance $Y$ from the axis of the
sphere. $X$ and $Y$ are in units of $R$. Solid squares: a Laponite
suspension similar to that described in the text; $G'$ is 1.5
times larger due to two months additional storage prior to use.
Here $R = 1.26$ cm and $v/R = 1.5$ s$^{-1}$. Crossed squares:
Carbopol, a non-aging yield-stress fluid \protect\cite{tcd06}; $R
= 1.96$ cm and $v/R = 0.57$ s$^{-1}$. Open squares: the initial
positions of the first seven of the particles from the Carbopol
experiment, shown to emphasize that lateral displacements are
significant in this case. Solid line: calculated displacements for
a Newtonian fluid with no wall effects; here the results do not
depend on $v$. The final position of the sphere is shown by the
dotted line. The inset shows the position of a particular tracer
particle at successive times (open circles) and the corresponding
locations of the sphere's center (solid circles). }
\end{figure}

The thinness of the fluidized layer for an aging suspension
contrasts markedly with the situation in non-aging yield-stress
fluids, for which the fluidized region extends approximately $R$
from the surface of the sphere \cite{bm97,btab85}. This is
expected to result in significant irreversible flow in the
material ahead of the sphere in the latter case, but not in the
former. We confirmed this by tracking the motion of 360 $\mu$m
glass beads suspended in the material and illuminated by a laser
sheet. Figure \ref{displacement} shows the displacement of an
initially horizontal line of tracer particles due to the slow
passage of the sphere in a Laponite suspension and in a non-aging
Carbopol gel \cite{tcd06} (here $F/F_c = 1.5$). The displacement
calculated for a Newtonian fluid \cite{pga00} is also shown. The
amplitude of the deformation is much smaller in the Laponite
suspension, and does not change significantly as the nominal shear
rate $v/R$ is varied from 1 s$^{-1}$ to 30 s$^{-1}$. The inset to
Fig. 4 shows the trajectory of a particle initially located about
$R/3$ from a vertical line through the center of the sphere in the
Laponite suspension. This particle moves horizontally to let the
sphere pass by then nearly returns to its initial position, but is
not displaced significantly in the vertical direction.

We have shown that the fluidized region around a moving sphere is
approximately two orders of magnitude thinner in a suspension
which exhibits rheological aging than in a non-aging material.
This behavior appears to be related to the shear-banding
instability observed in such materials in simple shear, but the
flow in the present case is more complex. Our results are relevant
to understanding the motion of objects in yield-stress materials.
While we studied the motion of a macroscopic sphere, similar
behavior may occur on much smaller scales around microscopic
particles diffusing within the suspension \cite{gvw05}, or even
around the colloidal particles that make up the suspension itself,
since the stresses involved are of the same order of magnitude.

We thank T. Toplak for his contributions. This research was
supported by NSERC of Canada. PC acknowledges receipt of a
visiting fellowship from the Center for Chemical Physics at UWO.

\vfill\eject

\begin{thebibliography}{99}

\bibitem{c05} P. Coussot, Rheometry of pastes, suspensions and granular
materials (Wiley, New York, 2005).

\bibitem{btab85} A.N. Beris, J.A. Tsamopoulos, R.C. Armstrong, and R.A.
Brown, J. Fluid Mech. \textbf{158}, 219 (1985).

\bibitem{tcd06} H. Tabuteau, P. Coussot, and J. R. de Bruyn, submitted to J. Rheol.

\bibitem{sollich} P. Sollich, F. Lequeux, P. H{\'e}braud, and M.
E. Cates, Phys. Rev. Lett. \textbf{78}, 2020 (1997).

\bibitem{d00} C. Derec, A. Ajdari, G. Ducouret, and F.
Lequeux, C. R. Acad. Sci. Paris IV \textbf{1}, 1115 (2000); L.
Ramos and L. Cipelletti, Phys. Rev. Lett. \textbf{87}, 245503
(2001); M. Cloitre, R. Borrega, F. Monti, and L. Leibler, Phys.
Rev. Lett. \textbf{90}, 068303 (2003).

\bibitem{c02}P. Coussot et al., Phys. Rev. Lett. \textbf{88}, 218301 (2002); F.
Varnik, L. Bocquet, J.-L. Barrat, and L. Berthier, Phys. Rev.
Lett. \textbf{90}, 095702 (2003).



\bibitem{acu95} D. D. Atapattu, R. P. Chhabra, and P. H. T. Uhlherr, J.
Non-Newtonian Fluid Mech. \textbf{59}, 245-265 (1995)

\bibitem{bm97} M. Beaulne and E. Mitsoulis, J. Non-Newtonian Fluid Mech.
\textbf{72}, 55 (1997).


\bibitem{f04} T. Ferroir, H. T. Huynh, X. Chateau, and P. Coussot,
Phys. Fluids \textbf{16}, 594-601 (2004); A. Khaldoun, E. Eiser,
G. H. Wegdam, and D. Bonn, Nature (U.K.) \textbf{437}, 635 (2005).

\bibitem{laponite} Southern Clay Products, Gonzalez, TX. http://www.laponite.com

\bibitem{bkhlm03} M. Bellour, A. Knaebel, J. L. Harden, F. Lequeux,
and J.-P. Munch, Phys. Rev. E \textbf{67}, 031405 (2003).

\bibitem{origin} In the inset to Fig. \ref{vsdeltarho}, $t=0$ corresponds to the
start of the experiment when the sample age is $t_w$. Elsewhere we
have removed data from the inertial regime by a small shift of the
time origin (0.8 s after the sphere enters the material in Fig.
\ref{vsdeltarho}); this shift does not affect our results.

\bibitem{c06} P. Coussot, H. Tabuteau, X. Chateau, L. Tocquer, and G. Ovarlez,
submitted to J. Rheol.

\bibitem{steady} There can be no truly steady
motion in a material in which the yield stress increases with age.
Our meaning here is simply that the measured velocity remains
constant over the duration of the experiment.

\bibitem{pmp96} F. Pignon, A. Magnin and J.M. Piau, J. Rheol. \textbf{40}, 573
(1996); P. Coussot, Q. D. Nguyen, H. T. Huynh, and D. Bonn, Phys.
Rev. Lett. \textbf{88}, 175501 (2002); D. Bonn, P. Coussot, H. T.
Huynh, F. Bertrand, and G. Debr{\'e}geas, Europhys. Lett.
\textbf{59}, 786 (2002).

\bibitem{da67} M. P. du Plessis and R. W. Ansley, J. Pipeline Div.
ASCE \textbf{2}, 1 (1967).

\bibitem{pga00} T. Papanastasiou, G. Georgiou, and A. Alexandrou,
Viscous Fluid Flow (CRC Press, Boca Raton, 1999).

\bibitem{gvw05} M. L. Gardel, M. T. Valentine, and D. A. Weitz, in
Microscale Diagnostic Analysis, edited by K. Breuer (Springer
Verlag, Berlin, 2005); T. A. Waigh, Rep. Prog. Phys. \textbf{68},
685 (2005).

\end{thebibliography}
\end{document}